# A transient thermal cloak experimentally realized through a rescaled diffusion equation with anisotropic thermal diffusivity

Yungui Ma[1], Lu Lan[1], Wei Jiang[1], Fei Sun[1,2] & Sailing He [1,2]

[1]*Centre for Optical and Electromagnetic Research, State Key Lab of Modern Optical Instrumentation, Zhejiang University, Hangzhou 310058, China*
[2]*Department of Electromagnetic Engineering, School of Electrical Engineering, Royal Institute of Technology, S-100 44 Stockholm, Swede*

**Abstract**

**Transformation optics originating from the invariance of Maxwell's equations under the coordinate mapping has enabled the design and demonstration of many fascinating electromagnetic devices that were unconceivable or deemed impossible before[1-11], and has greatly contributed to the advancement of modern electromagnetism and related researches assisted with the development of metamaterials[12-15]. This technique has been extended to apply to other partial differential equations governing different waves[16-23] or flux[24-28], and has produced various novel functional devices such as acoustic cloaks[20-23] and Schrödinger's 'hat'[19]. In the present work we applied the coordinate transformation to the time-dependent heat diffusion equation[24-28] and achieved the manipulation of the heat flux by predefined diffusion paths. In the experiment we demonstrated a transient thermal cloaking device engineered with thermal metamaterials and successfully hid a centimeter sized strong 'scatter' (thermal disturber), i.e., a vacuum cavity. To facilitate reliable fabrication we adopted the rescaled thermal diffusion equation for various ingredient materials with nearly constant product of the density and heat capacity, and took the anisotropic thermal diffusivities as the key parameters for the design. Our results unambiguously show the practical possibility to implement the complex transformed thermal media with high accuracy and acquire some unprecedented thermodynamic functions, which we believe will help to broaden the current research and pave a new way to manipulate heat for novel device applications.**

Corresponding e-mail: sailing@kth.se

Transformation optics (TO), a very powerful mathematical tool, has regained great research interests in the community since the pioneering works on electromagnetic (EM) cloaking independently done by J. Pendry[2] and U. Leonhardt[3]. Under the invariance of Maxwell's equations, the coordinate modification can transform virtual geometries into concrete physical EM media usually parameterized by anisotropic constants, enabling the manipulation of lights almost as arbitrarily as we desire[29]. Excited by the tremendous application potential great endeavors have been made to push the EM cloaking toward the practice in different spectra or dimensions[2-13]. The TO technique has also led to the creation of many other important EM devices with functionalities deemed impossible or unconceivable before such as omnidirectional retroflection lens[14], wave rotator[30], or illusionary optical devices[31]. In general this coordinate operation can apply to different partial differential equations (PDEs) governing the behaviors of other types of waves. As a result transformed devices able to control physical waves or flux such as acoustic cloak[20-23], Schrödinger's 'hat'[19] and thermal cloak[24-28] have been proposed showing important scientific and application potentials.

In this work we implement the concept of transformation thermodynamics on the general time-dependent heat equation and demonstrate a thermal device able to transiently segregate a region from the surrounding temperature field without disturbing the external heat flux, thus rendering a transient thermal cloaking. In the literature[24-26] thermal cloaks have been previously proposed but mostly by applying coordinate transformation to the thermostatic heat conduction equation, i.e., $\nabla \cdot \kappa \nabla u = 0$ with $\kappa$ as the thermal conductivity and $u$ the temperature field. This limits the transformed devices to work only at a thermostatic state. A beautiful experimental work has been recently reported using engineered thermal metamaterials to elaborate a stationary thermal cloak and other transformed devices as well[27]. For many practical applications a device that can work at a transient state is more favorable or necessary to control the dynamic heat flux for specific purposes such as heat shield[31] or energy harvesting[33]. This requires dealing with the general time-dependent thermal diffusion equation

$$\rho c \frac{\partial u}{\partial t} = \nabla \cdot (\kappa \nabla u), \qquad (1)$$

where $\rho$ and $c$ are the density and the specific heat capacity of the medium, respectively. This equation describes the heat diffusion in a solid out of the source region. Applying the invariance role of the thermal PDE under a coordinate mapping from the virtual $(x, y)$ into the physical $(x', y')$ spaces, Guenneau et al recently derived the transformed medium that equally satisfied the heat equation[28]

$$\rho' c' \frac{\partial u}{\partial t} = \nabla' \cdot (\boldsymbol{\kappa}' \nabla' u) \quad (2)$$

with the new material parameters $\rho'c' = \rho c/\det(\boldsymbol{J})$ and $\boldsymbol{\kappa}' = \boldsymbol{J}\kappa\boldsymbol{J}^T/\det(\boldsymbol{J})$, where $\boldsymbol{J}$ is the Jacobean matrix expressed by $\partial(x', y')/\partial(x, y)$. It is clear that a device of transient response needs to take the density and heat capacity into account in addition to the anisotropic thermal conductivities, thus making it more challenging to implement. In this work we managed to realize these complex parametric requirements and fabricated such a device exhibiting a prominent cloaking behavior for transient heat flux. This is enabled by the abundance of natural materials with a wide range of thermal conductivity from nearly zero to a value> 1000, which is hardly available in the design of an EM device[4,14]. Our work unambiguously proves the feasibility to extend the EM TO devices into the thermal area, which may pave a unique way to steer the dynamic heat flux for novel device applications.

**Results**

**Design the samples**

For the simplicity of implementation we discuss the transformation technique and its application in thermal devices in two dimensions (2D). Figures 1a and 1b show the schematic diagram to create a thermal cloak using the transformation technique. Similar to the design of an EM cloak[2-4], transformation allows us to expand a point in the virtual space (Fig. 1a) into a thermally invisible cavity in the physical space (white region in Fig. 1b) enclosed by an annular cloak (interior radius $R_1$ and exterior radius $R_2$). The heat flux is re-guided to flow around the cavity and restore their original diffusion paths outside the cloaking shell. As analyzed in Ref. [28], for an ideal thermal cloak, the heat energy takes an infinite time to pass through the innermost cloaking shell (which has nearly zero radial conductivity) before reaching the center cavity, thus offering a thermal protection. However, in practice we need to use finite meshes in simulation or thin layers in experiment to approximate the continuous device with finite conductivity values. As a result thermal energy will slightly diffuse into the inner cavity region and raise the temperature as the time elapses although this process is slower than that in the surrounding background, thus only offering a temporal heat protection. The same thing happens for a thermostatic cloak with the temperature inside the cloaking region dependent on the spatial position where the transformation starts as shown in Fig. 1a[26,27]. The examples to show this behavior will be given later in discussing the simulation results in Figs. 2a-2l. Here we would like to stress that a thermal cloak mainly acts as a flux-guiding shell to suppress the influence of a 'scatter' (thermal disturbance) in addition to the temporal thermal protection. One may introduce a thin thermally insulating layer enclosing the cavity to improve the protection, as later indicated by our experiment in cloaking a vacuum cavity.

In our design and later fabrication a simple linear coordinate mapping is chose to construct the material profiles of the thermal device, i.e.,

$$r = \frac{R_2-R_0}{R_2-R_1}r' - \frac{R_2(R_1-R_0)}{R_2-R_1}, \quad (3)$$

where $r$ ($r'$) is the radius in the virtual (physical) space. Here we assume the expansion as shown in Fig. 1a starts from a finite small region of radius $R_0$ (=0.005$R_1$) rather than exactly from a point. This approximation on the initial condition leads to negligible influence on the device performance but does reduce the demand on material parameters. In this work our device is designed with $R_1$ = 13 mm and $R_2$ = 20 mm. Supplementary Fig. S1 shows the transient cloaking behavior for the ideal case simulated by COMSOL. To implement such a device one needs to discretize the continuous device and approximate it by a number of homogeneous anisotropic concentric layers. In the limits of thermostatic circuitry, each anisotropic sub-layer can be approximated by a number of isotropic ingredient layers[26-28]. It's clear that at least four different ingredient layers are required to solve Eq. (2) which has three unknowns, i.e., $\rho'c'$ and the two principal values of anisotropic $\boldsymbol{\kappa}'$. By the materials used in our experiment, as listed in supplementary Table S1, we can mathematically fulfill the parametric requirements but only for a few sub-layers. In order to improve the fabrication efficiency here we modify the original equation by transferring $\rho'c'$ to the right side, i.e.,

$$\frac{\partial u}{\partial t} = \nabla' \cdot (\boldsymbol{\kappa}'/\rho'c'\nabla'u). \quad (4)$$

We see the rescaled Eq. (4) is effectively equivalent to Eq. (2) if the condition $\nabla'(1/\rho'c') \approx 0$ is satisfied, which leads to $\nabla'\cdot(\boldsymbol{\kappa}'/\rho'c'\nabla'u) = 1/\rho'c'\nabla'\cdot(\boldsymbol{\kappa}'\nabla'u)+\boldsymbol{\kappa}'\nabla'u \cdot\nabla'(1/\rho'c') \approx 1/\rho'c'\nabla'\cdot(\boldsymbol{\kappa}'\nabla'u)$. This sets up a precondition to use Eq. (4) that the implemented device should have a nearly constant product of density and heat capacity across the sub-layers. With this condition satisfied we now only need to implement the anisotropic thermal diffusivity $\boldsymbol{\alpha}'(= \boldsymbol{\kappa}'/\rho'c')$, which offers more freedom to choose $\boldsymbol{\kappa}'$. The transformed diffusivity profile in a cylindrical coordinate is calculated by

$$\boldsymbol{\alpha}' = diag\left\{\frac{r}{r'}\frac{dr'}{dr}, \frac{r'}{r}\frac{dr}{dr'}, \frac{r}{r'}\frac{dr}{dr'}\right\}\alpha_0, \quad (5)$$

where $\alpha_0$ is the diffusivity of the background. To implement this parametric profile defined by Eqs. (3) and (5), we discretized the device into seven sub-layers with each layer having an equal thickness of 1 mm and engineered them with many thin ingredient sheets (their properties are given in supplementary Table S1). The averaged parameters for each sub-layer are calculated by the following formula

$$\overline{\rho'c'} = \sum_{i=1}^{N} f_i\rho_ic_i, \quad \overline{\alpha'_r} = (\overline{\rho'c'}\sum_{i=1}^{N}\frac{f_i}{\kappa_i})^{-1}, \quad \overline{\alpha'_t} = \sum_{i=1}^{N} f_i\kappa_i/\overline{\rho'c'}, \quad (6)$$

where $\rho_i$, $c_i$ and $\kappa_i$ are, respectively, the density, heat capacity and conductivity of the $i^{th}$ ingredient sheet whose volume ratio $f_i$ satisfies $\sum_{i=1}^{N} f_i \approx 1$ (note this approximation was due to the fact that each ingredient layer has finite thickness and their summation may not be exactly unit). $\overline{\alpha'_r}$ and $\overline{\alpha'_t}$ are the averaged radial and tangential thermal diffusivities, respectively. Here we need to point out EM metamaterials are usually engineered with unit-cell sizes far smaller than the operating wavelength to validate the usage of effective medium theory[34]. For the thermal case, the ingredient materials composing the sub-layers should be selected with thicknesses as small as possible in order to have a good accuracy in using the average equations (6) based on the analogy of a static thermal circuitry. Experimentally different ingredient sheets of thicknesses varying from 0.025 to 0.2 mm have been used to fabricate the 7 mm thick cloaking shell which has a total sheet number of 79. The details on the structural parameters and ingredient materials for each sub-layer are listed in Table 1, where we assume the background has an isotropic diffusivity of $\alpha_0$ = 34 mm$^2$/s. It is worth mentioning that the metals and pyrolytic graphite (of the largest anisotropic conductivity) used here are of good thermal stability in properties within the interested temperatures from zero to ~300°C[35], which is very important for the design and application. From Table 1, we see that the designed layers have roughly equal $\overline{\rho' c'}$ around 2.1 J/cm$^3$/K except for the first layer (i.e., the innermost layer), which guarantees the rationality in using Eq. (4). The parametric discrepancy of the first sub-layer does affect the overall cloaking performance but not seriously as numerically shown in Figs. 2a-2l. This may be understandable because the sub-layer is very thin (1 mm) and the diffusivity is more decisive.

**Simulation results**

Figures 2a-2d show four snapshots of the temperature field captured at time $t$ = 0, 3, 21 and 90 s, respectively, for the device defined by the parameters given in Table 1. In simulation we used the normalized constant temperature as the left ($T_1$ = 1) and right ($T_2$ = 0) boundaries and the thermal insulation for the top and bottom boundaries. For the first example, the background and the cloaking region are filled by the same material of $\rho c$ = 2.1 J/cm$^3$/K and $\kappa$ = 34 × 2.1 = 71.4 W/m/K. As indicated by the white isothermal lines, the heat flux flows around the cloaked region and restores their straight diffusion trajectories (orthogonal to the isothermal lines) after passing it. The temperature field outside the cloak is only slightly disturbed as if no disturber exists in the flux paths. We also see the slow increment of the temperature inside the cloaked region as time elapses with the reason addressed earlier. At $t$ = 90 s, the thermal system almost reaches the equilibrium state and the temperature field mimics the ideal case (supplementary Fig. S1). These snapshots and their close match with the ideal case justify the approximations made to rescale the thermal equation and homogenize the continuous profile by multiple layers in the process of implementation. In the experiment we use brass (Cu$_{70}$Zn$_{30}$) as the background material whose $\rho c$ and $\kappa$ are both 1.56 times larger than those used in the above simulation (i.e., they have the same diffusivity). This reduced background will lead to the thermal impedance ($\propto 1/\kappa$ in the static limit) mismatch at the device's exterior boundary and degrade the cloaking performance by causing a weak 'shadow' after the cloak, as shown in Figs. 2e-2l. It is similar to the rescaling measure adopted in the EM cloaks to acquire a constant tangential permeability[4,14,36]. We take this sacrifice to conduct the experiment with available materials and in a reasonable measuring period. Figures 2i-2l show the snapshots of the temperature field with the cloaking region replaced by a vacuum cavity, an absolute 'scatter', which will act as similarly as the perfectly electric conducting (PEC) metal used in an EM cloak[4,6]. In this case no energy will diffuse into the center region. We see the temperature field has almost no difference compared with those shown in the second row, indicating the robust cloaking performance of our designed devices, i.e., insensitive to the materials put inside the cloaking region.

**Measurement results**

Figure 3a gives a photo of a fabricated thermal cloak consisting of 79 ingredient sheets. The quality of the device, in particular the influence of the thermal interface resistance between sheets, needs to be experimentally examined. Figure 3b gives the anisotropic diffusivity values of each sub-layer with the lines calculated by Eq. (5) and the symbols from Table 1. We managed to implement the device with the effective parameters (for effective realization) as close as possible to the theoretical values by carefully selecting the ingredient sheets (species and thickness) that simultaneously well satisfy $ρ'c'$ = *constant* for most layers. This is generally accomplished by using multiple ingredient materials for each sheet. Figure 3c shows a schematic of our experimental setup conducted in a vacuum environment. Here we used an infrared camera to capture the temperature field of the sample through Planck thermal emission. More details about sample fabrication and measurement could be found in the Method section. In our experiment we fabricated two cloak samples with one having a brass center and one having a vacuum cavity in the cloaking region. For reference we also measured the temperature field of a same-sized bare brass.

Figure 4a gives a photograph of a rectangular bare brass and Figs 4b-4d show its transient conduction behavior with the temperature snapshots respectively captured at 2, 6, and 15 min after switching on the power of the heater. The isothermal lines are plotted in the white curves. An animation movie with more photos is given in supplementary Movie S1. A nearly ideal planar temperature front, indicating a straight heat flux, is observed for the empty background. Note that our measurement time period is far larger than that used in simulation, as the constant-power source used here needs a long time (~ 13 min) to reach the thermal equilibrium for itself. However, this difference will not disturb our evaluation on the samples' performances referring to the simulation results. Figure 4e gives a picture of our first sample for cloaking a brass disk. The two small holes in the disk are left in the manufacture. Figures 4f-4h show the snapshots of the temperature field captured at *t* = 2, 6, and 15 min, respectively. More photos for an animation movie are given in supplementary Movie S2. From the isothermal lines we see that the heat flux (orthogonal to the isothermals) is guided to flow around the center disk and diffuse forward with little disturbance as compared to the top figures for a bare background. The brass disk has a nearly uniform temperature profile rising slowly with the absolute values much smaller than those in the left surrounding medium. This transient behavior is consistent with the numerical predictions shown in Figs. 2e-2h.

Figure 4i gives a photograph of our second cloak sample with a cavity hole (radius = 9 mm) in the cloaking center. Note here we have used a 4 mm thick brass ring to support the cloak shell. Figures 4j-4l show the snapshots of the temperature field captured at 2, 6, and 15 min, respectively. An animation picture is given in supplementary Movie S3. Assisted by the isothermal line these figures show that the heat flux is guided around the cavity and restores their diffusion direction after the cloak. The strong 'scatter' (thermal disturbance) has a slight influence on the heat flux and the temperature field as indicated by slightly more distorted isothermal lines after the cloak compared with Figs. 4g and 4h with a solid in the cloaking region. The experimental results agree well with the simulations shown in Figs. 2i-2l. We should emphasize that the distortion of the isothermal lines near the sample is mainly caused by the thermal impedance mismatch between the device and background as addressed in our previous analysis. We believe the cloaking performance can be further enhanced with a background of matched impedance as shown in Figs. 2a-2d.

**Discussions**

In this work we have implemented a coordinate-transformed thermal device and demonstrated the transient thermal cloaking behavior. To facilitate the fabrication we adopted the rescaled thermal diffusion equation and took the anisotropic thermal diffusivity (instead of the thermal conductivity in all the previous literature) as the key parameter for the design. This treatment is fundamentally different from a thermostatic case where only conductivity is considered and is practically validated by satisfying the prerequisite on the product of the density and heat capacity and thus being intrinsically consistent with the general time-dependent thermal equation. We would like to point out that the rescaling equation is also different from the previous approximation made in Ref. 28 to rescale the conductivities in order to homogenize the heat equation, which is valid only near *r'* = $R_1$ according to their transformation algorithm. The strict requirements on the material parameters in our design are deliberately fulfilled by properly choosing multiple ingredient layers rather than two alternating layers used for the static device[26,27]. In the process machining and assembling the devices must be carefully conducted, otherwise defects like thermal interface resistance will be induced which may easily ruin the final parametric profile and consequently the sample performance. Our experimental results explicitly show the possibility to accomplish such a complex sample with engineered thermal metamaterials. Differing from the EM metamaterials designed for transformed wave devices, which are usually highly dispersive and of narrow band response[4], thermal metamaterials and the implemented devices can be fabricated by natural materials and thus are more promising and robust for practice (the temperature issue could be solved by controlling the ingredient species and the temperature window for application).

Transformation optics has brought about a major advancement of modern electromagnetic research by producing many fascinating EM devices that were previously unconceivable or deemed impossible. We may also believe that transformation thermodynamics can also lead to some revolutionary thermal technologies or devices through the precise manipulation of heat flux. Applications can be envisioned such as transformed devices to improve the efficiencies of energy harvesting for thermoelectrics by creating low thermal and high electric conductivity profiles[33] or to fulfill a specific heat flux channel for a thermal circuit[37]. Transformation thermodynamics may also provide unique ways to solve the problem in heat management encountered by a chip designer[32], e.g., constructing a specific substrate or embedding a medium to guide away the heat. We may also have a perspective to integrate the transformation optics and thermodynamics in a single device to achieve simultaneous manipulation of both light and heat, excited for thermal-optical applications such as thermal photovoltaic devices[38].

**Methods**

**Simulation.** In this work we discuss a thermal cloaking device in 2D. Experimentally it is satisfied by measuring a relatively thin sample in a vacuum environment. Our samples are designed using the rescaled Eq. (4) under the precondition $\rho'c'$ = *constant*. The anisotropic diffusivity and this prerequisite are experimentally satisfied by properly choosing and assembling the ingredient sheets. Simulation is done by COMSOL to examine the approximations and predict the cloaking performance. The cloaking behavior for an ideal thermal cloak using Eqs. (2) and (3) is shown in supplementary Fig. S1. In the article we give the simulation results for the devices with parameters experimentally implemented. In our simulation the influence of other heat diffusion channels such as the thermal radiation and convection are neglected considering the practical condition, i.e., a 5 mm thick sample measured in a vacuum environment.

**Fabrication.** Great care has been taken to fabricate the samples. Our cloak shells consist of two equal halves that are separately manufactured and merged together later. In fabrication we used the raw materials listed in supplementary Table S1 to construct our device. These ingredient sheets are carefully selected with the help of numerical simulation to meet the anisotropic diffusivity profile defined by Eq. (5) and the requirement for a constant $\rho'c'$. For each sub-layer we mix the different ingredient sheets as homogeneously as possible by repeating their basic combination. Our implemented devices have a total of 79 layers of ingredient sheets. Compressing stress must be suitably applied in assembling these layers in order to have a gapless sample without deformation. We used a 15 cm × 10 cm rectangular brass as the background material, which has 1.56 times larger $\rho c$ and $\kappa$ than the thermal-impedance matched case (the matched ground has $\rho c$ = 2.1 J/cm³/K and $\kappa$ = 34 × 2.1 = 71.4 W/m/K). This reduction does sacrifice the cloaking performance as numerically shown in Figs 2e-2l.

**Measurement**. Our measurement was conducted in a vacuum chamber with an infrared window on the top to measure the radiated infrared light that is associated to the temperature field of the sample by Planck thermal emission. Rather than a constant temperature source as used in simulation, we used a 40 W strip heater glued on the left edge of the brass by silicon grease to produce the heat and submerged the right edge in an ice-water sink to attenuate the heat. Evolution of the temperature field was captured by an infrared camera Mikron 7500L at a time interval of 1.5 s between two successive shots. The vacuum pressure in the measurement is maintained below 2 Pa. The thermal emission efficiency of the sample was improved by spraying a ∼ 20 μm thick blackbody paint (Okitsumo) on the top surface of the sample, whose influence on the thermal diffusion of the sample could be neglected. Our setup allows the measurement of a 2D thermal diffusion behavior, as exactly required by our design. A photograph of our measurement setup is given in supplementary Fig. S2.

**Acknowledgments**

The authors are grateful to the partial supports from NSFCs 61271085, 60990322 and 91130004, the National High Technology Research and Development Program (863 Program) of China (No. 2012AA030402), NSF of Zhejiang Province (LY12F05005), the Program of Zhejiang Leading Team of Science and Technology Innovation, NCET, MOE SRFDP of China, and Swedish VR grant (# 621-2011-4620) and AOARD.

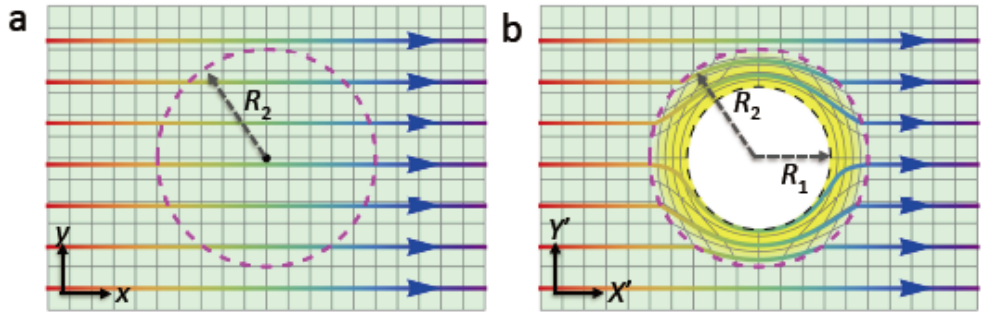

**FIGURE 1** Schematic diagram of coordinate transformation for a thermal cloak. (a) The empty virtual space $(x, y)$ and (b) the transformed physics space $(x', y')$. A simple linear coordinate transformation expands the central point in (a) into a white cavity enclosed by an annular cloaking shell (interior radius $R_1$ and exterior radius $R_2$). The colorful lines in the figures represent the diffusion trajectories of the heat flux.

| No. | Ti (%) | Al (%) | Cu (%) | p-G (%) | Ka (%) | $\rho'c'$ (J/cm³/K) | $\alpha_r'$ (mm²/s) | $\alpha_t'$ (mm²/s) |
|---|---|---|---|---|---|---|---|---|
| 1 | - | - | 0.1 | 0.84 | 0.05 | 1.22 | 1.71 | 623 |
| 2 | - | - | 0.50 | 0.49 | 0.012 | 2.19 | 3.4 | 277 |
| 3 | - | 0.7 | - | 0.28 | 0.007 | 2.00 | 6.16 | 199 |
| 4 | 0.7 | - | - | 0.35 | - | 1.94 | 7.9 | 151 |
| 5 | 0.6 | 0.2 | - | 0.25 | - | 2.08 | 9.0 | 121 |
| 6 | 0.5 | 0.4 | - | 0.18 | - | 2.19 | 10.7 | 105 |
| 7 | 0.43 | 0.4 | - | 0.14 | - | 2.22 | 11.3 | 99 |

**TABLE 1** Material parameters of thermal cloaking device. Our cloaking device is composed of seven homogeneous sub-layers. We used five different ingredient sheets to make the sub-layer, namely, titanium (Ti), aluminum (Al), copper (Cu), pyrolytic graphite (p-G) and kapton (Ka). The properties of these materials are given in supplementary Table S1. The volume percentage of each ingredient in each sub-layer is calculated by Eqs. (5) and (6). The right three columns list the averaged product ($\rho'c'$) of the density and heat capacity, radial ($\alpha_r'$) and tangential ($\alpha_t'$) thermal conductivities designed for each sub-layer.

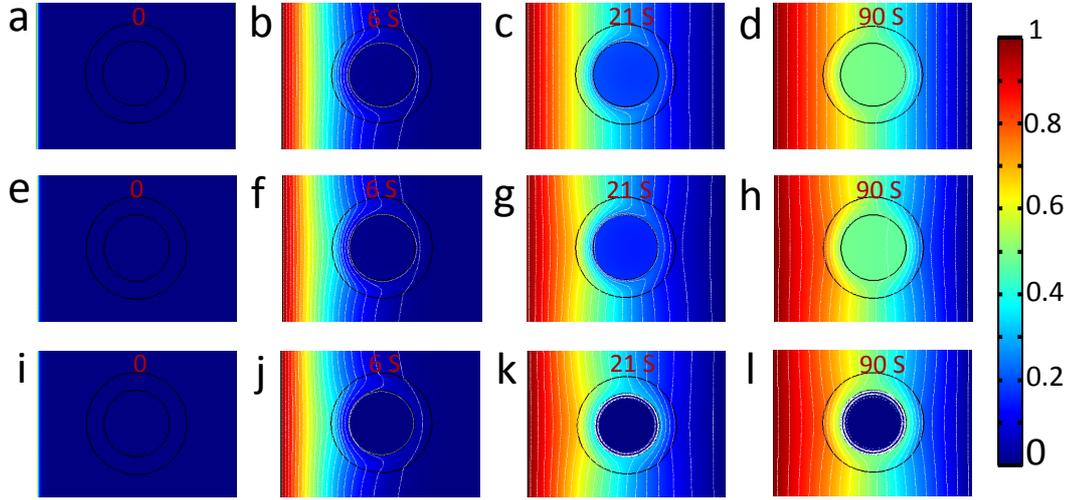

**FIGURE 2** Simulated cloaking behavior. The first to fourth columns list the snapshots of the normalized temperature distribution captured at $t$ = 0, 6, 21 and 90 s, respectively. (a)-(d) For the case with an impedance-matched background ($\rho c$ = 2.1 J/cm$^3$/K and $\kappa$ = 71.4 W/m/K) with the cloaking region filled by the same material as the background. The second (e)-(h) and the third (i)-(l) rows are for the cases using the impedance-mismatched background of brass (Cu$_{70}$Zn$_{30}$, $\rho c$ = 2.1 × 1.56 = 3.28 J/cm$^3$/K and $\kappa$ = 41.4 × 1.56 = 111 W/m/K) with the cloaking region occupied by a brass disk and a vacuum cavity, respectively. In simulation, heat flux originates from the left edge ($T_1$ = 1) and diffuses toward the right edge ($T_2$ = 0). The top and bottom edges are thermally isolated boundaries. The isothermal lines (orthogonal to the heat flux trajectories) are plotted in the white curves.

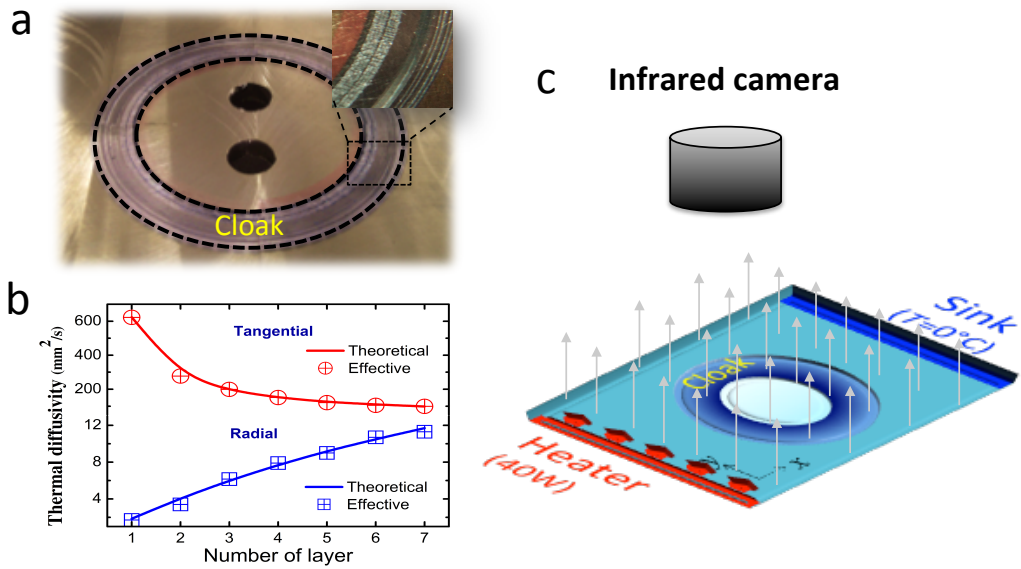

**FIGURE 3** Device and measurement. (a) A photo of a fabricated cloaking device. The cloak shell is highlighted by two dashed lines. The inset in (a) gives a zoon-in figure of part of the cloak shell. (b) The tangential (top) and radial (bottom) thermal diffusivities of different sub-layers. The solid lines represent the theoretical values calculated by Eq.(5) and the symbols represent the effective values designed for each sub-layer using Eq. (6). (c) Schematic diagram of our measurement setup in vacuum. A 40 W strip heater source is glued to the left edge of the brass and an ice-water sink is used at the right edge to induce the heat flux. The other boundaries in vacuum mimic thermal insulation conditions, consistent with the simulation. Evolution of the temperature field is captured by an infrared camera at every time interval of 1.5 s.

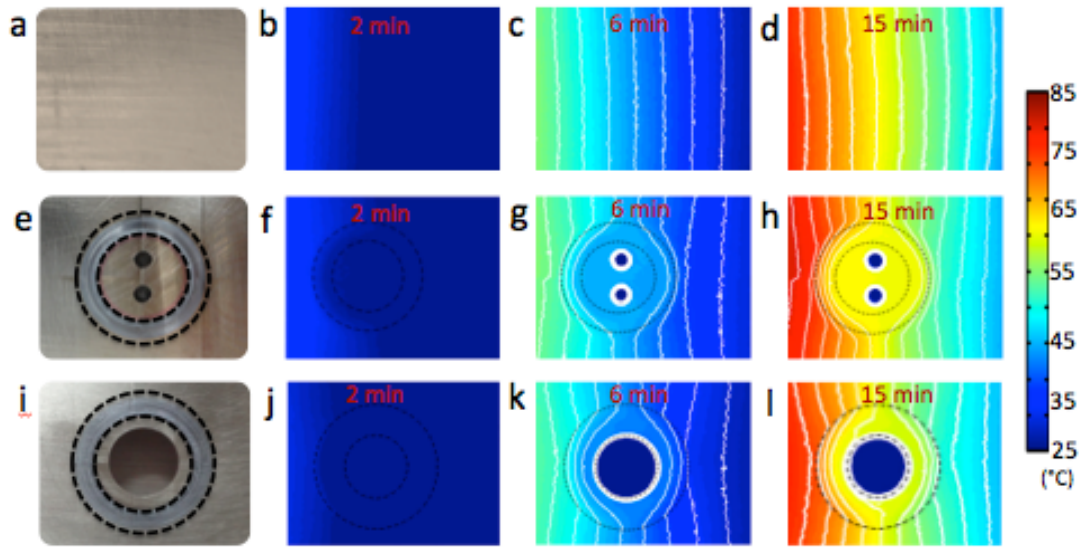

**FIGURE 4** Samples and measured temperature field. (a) A top view of a rectangular bare brass. (b)-(d) Corresponding snapshots of the temperature field captured at $t$ = 2, 6, & 15 min, respectively, for the bare brass background. (e) A top view of the cloaking device with a brass disk in the cloaking region. The two small holes are left in the manufacture. (f)-(h) Corresponding snapshots of the temperature field captured at $t$ = 2, 6, & 15 min respectively for this sample. (i) A top view of the cloaking device with a vacuum cavity in the cloaking region. (j)-(l) Corresponding snapshots of the temperature field captured at $t$ = 2, 6, & 15 min respectively for this sample. The isothermal lines (orthogonal to the heat flux trajectories) are drawn in white. Animation figures to show the evolution of the temperature field are given in supplementary Movies S1-S3.